\documentclass{article}

\usepackage{arxiv}

\usepackage[utf8]{inputenc} 
\usepackage[T1]{fontenc}    
\usepackage{hyperref}       
\usepackage{url}            
\usepackage{booktabs}       
\usepackage{amsfonts}       
\usepackage{nicefrac}       
\usepackage{microtype}      
\usepackage{lipsum}		

\usepackage{graphicx}
\usepackage{caption}
\usepackage{subcaption}
\usepackage{lscape}
\usepackage{float}

\title{A Benchmark API Call Dataset for Windows PE Malware Classification}

\author{
	Ferhat Ozgur Catak \\
	Cyber Security Institute\\
	TUBITAK-BILGEM\\
	Kocaeli, Turkey \\
	\texttt{f.ozgur.catak@gmail.com} \\
	\And
	Ahmet Faruk Yazi \\
	Cyber Security Engineering Department \\
	Istanbul Sehir University\\
	Istanbul Turkey \\
	\texttt{ahmetyazi@std.sehir.edu.tr} \\
}

\begin{document}

\maketitle

\begin{abstract}
The use of operating system API calls is a promising task in detecting PE-type malware in the Windows operating system. This task is officially defined as running malware in an isolated sandbox environment, recording the Windows operating system's API calls and sequentially analyzing these calls. Here, we have analyzed 7107 different malicious software belonging to various families such as virus, backdoor, trojan in an isolated sandbox environment and transformed these analysis results into a format where different classification algorithms and methods can be used. First, we'll explain how we got the malware, and then we'll explain how we've got this software bundled into families. Finally, we will describe how to perform malware classification tasks using different computational methods for the researchers who will use the data set.

\end{abstract}





\section{Introduction}
Nowadays, the use of computers in daily life is becoming widespread, and as a result, computer attackers are attacking computers using different methods, or they use these computers as weapons. Although computers become more secure with each new operating system version or update, attackers can bypass these security components using different methods. The most common scenario of security component bypass methods is that the malware changes its source code and behaviour on each infected computer \cite{7952603}. All of the methods used by analysts to detect malicious software is called malware analysis. Malware analysis is a broad term and includes many stages. These stages involve examining the suspicious software's contents without running the software and then running the software in an isolated environment, exploring the domain name system (DNS) resolution requests, recording the registry reads/writes, file accesses and the application programming interface (API) calls. 

These malware act for a specific purpose. We know that they are used for many different purposes, such as preventing a system from working, gaining unauthorized access to a network, obtaining personal data. For such purposes, many other platforms such as servers, personal computers, mobile phones, and cameras are targeted. Today, the number of platforms that have become a target is increasing. Consequently, malicious software developed for these platforms is also quite different. In particular, in the first 4 months of 2018, 40,000,000 is considered the face of the danger of new malware.

Nowadays, there is a considerable amount of time spent to protect from this software and significant budget expenditures. In order to protect against malicious software, many products are produced both commercially and academically. There is a severe struggle between attackers creating malware and the parties trying to identify this software. As a result of this situation, it increases the competencies and capabilities of both sides. Metamorphic malware is the result of this struggle.

Metamorphic malware is the most advanced members of the malware family. Using different methods, this malicious software can make their structures change continuously by making changes in their source codes. In this way, they change the code signatures. Besides, this software may have the ability to recognize the environment and store their harmful actions by counter-analysis actions in the environments created for malware analysis \cite{8358312}. Metamorphic malware is difficult to detect and classify as they have such capabilities.

Considering the development of malware, it is observed that they underwent a structurally perfect evolution. But there is one constant characteristic in each phase. These features are designed to benefit unpleasantly. That is, they have harmful behaviour. All malware should perform some actions to achieve its goals. Assuming that malware is on a computer running the Windows operating system, this malware needs to use some of the operating system's services. The entire set of requests to get these services (Windows API calls) creates malicious behaviour. Malicious software detection and classification can be performed if such malicious behaviour is well analyzed.

Detection of malicious software includes many design issues that need to be addressed, such as incorrect jump opcodes in the assembly codes, hidden content in the \textit{.text} block in portable executable (PE) file, and encrypted content. This study collected the current malware and its variants, such as WannaCry and Zeus, especially on the Github website. We obtained the family classes from the VirusTotal site by finding the hash values of each malware. Finally, all the behaviours were recorded by running them in a Cuckoo sandbox environment. We argue that almost all malware change their behaviour using a variety of methods. Although they change their behaviour, malicious software has a target and have a pattern of paths to achieve this goal. Furthermore, the malware makes unnecessary API calls during the behaviour change, and it can be detected by a model to be trained by analysts because the pattern is the same.

Malware analysis may be defined as the branch of cybersecurity, consisting of two phases: (1) static analysis, (2) dynamic analysis of suspicious files. The static analysis can broadly be defined as examining the executable file without viewing the actual instructions by executing an isolated environment. A well-known example of static analysis is MD5 checksums, recognition by antivirus detection tools, finding strings. Dynamic analysis refers to actual run malware to understand its functionality, observe its behaviour, identify technical indicators. The most important part of the behavioural records is API call sequences. Most studies in the field of dynamic malware analysis have only focused on classification algorithms. The fundamental problem with this research is that there are no benchmark datasets to check the efficiency of the proposed models \cite{7166115}.

This study seeks to obtain data that will help to address these research gaps. This study's specific objective is to build a benchmark dataset for Windows operating system API calls of various malware. This is the first study to undertake metamorphic malware to make sequential API calls. It is hoped that this research will contribute to a deeper understanding of how metamorphic malware change their behaviour (i.e. API calls) by adding meaningless opcodes with their own dissembler/assembler parts. 

We shared our data set over GitHub site \footnote{\url{https://github.com/ocatak/malware_api_class}}. We believe that this dataset can be used by researchers who conduct studies on behaviour-based malware analysis. The datasets is used on several malware analysis research: LSTM based detentions, \cite{10.7717/peerj-cs.285,8806571} CNN based modelling \cite{10.7717/peerj-cs.346}, the traditional approach \cite{asrafi2020comparing,city24815,tariq2020review}.


\section{Methods}
The dataset contains raw data regarding the cuckoo sandbox based known malware execution and VirusTotal based classification of files using their MD5 signatures.
\subsection{Windows API Calls}
The Windows API is an interface for developing applications on the Windows operating system. Application developers can communicate with your operating system using the Windows APIs. Therefore, the operating system offers many services as an API. A Windows application needs to use the APIs to use a function provided by the operating system. The use of these functions is defined as the API call. An application makes API calls many times during its execution. For example, when an application is requested to create a file, it must call the \textit{CreateFileA} API \cite{6890868}. API calls made by an application on the system can show the behaviour of this application. For this reason, API calls are often used in dynamic malware analysis. The primary entries of the data set used in this study are API calls made by malware on the operating system.

\subsection{Cuckoo Sandbox}

You can check any suspicious file in a few minutes with Cuckoo. It provides a detailed report showing the file's behaviour is executed in an isolated and realistic environment. Nowadays, it is not enough to detect and remove the effects of malware: it is vital to understand the context, motivations and how they work to understand a violation's purposes. Cuckoo Sandbox is free software that automates analyzing malicious files under Windows, OS X, Linux and Android. Cuckoo Sandbox is an advanced, highly modular and open-source automated malware analysis system with endless application possibilities.

In computer security, sandboxing is a security mechanism used to separate running programs. Usually, sandboxing is used for unconfirmed applications from third parties, suppliers, untrusted users, and untrusted websites. The Cuckoo Sandbox system has two primary components. The first component is the management machine, where the analysis of malware is started, the results are written to the database, and the web service is provided for the users. The second component is the analysis machines to run malicious software. Analysis machines can be virtual or physical machines \cite{8262998}.

\subsection{VirusTotal}
Virus Total is a free service that allows you to analyze files or URL addresses online. Many antivirus application engines and website scanners are used for analysis. Files considered to be harmful are analyzed individually in antivirus application engines. Each antivirus application engine creates an analysis report for the suspicious file \cite{7921994}. 

The same analysis case is valid for URLs to be analyzed. The VirusTotal service includes an extensive set of analyzes. In this way, a new scan can be performed, and previous analysis information can be obtained. Virus Total offers a service interface (VirusTotal Public API v2.0) to provide results without using a browser, as well as through a web browser. With this interface, files / URL addresses can be analyzed automatically.

Virus Total Public API provides the results of the analysis as a JSON object. The results of each antivirus application engine and web browser analysis are obtained separately.

\subsection{Dataset Creation}
The data set, as presented herein, has a straightforward structure. Our dataset is provided as comma-separated values (CSV) files to enhance interoperability, and no specific software or library is required to read them. The data were collected with \textit{git} command-line utility from various GitHub pages. Each row in this data set is an ordered sequence of Windows operating system API calls that belong to an analysis in the cuckoo sandbox environment.

The following steps were followed when creating the dataset.

\begin{enumerate}
	\item \textbf{Preparation of Cuckoo Sandbox Environment}: TThe Ubuntu operating system was installed on the analysis machine. Then the Cuckoo Sandbox application has been installed. The analysis machine is run as a virtual server, where malware will be run and analyzed. Windows operating system is installed on this server. The firewall has been turned off, and operating system updates have not been applied to prevent any obstacles during malicious software operation.
	\item \textbf{Analysis of malware}: More than 20,000 malware were run in Cuckoo Sandbox one at a time. The application has written the analysis information of each malware into the MongoDB database. From this analysis information, the behaviour data of the malware on the analysis machine were obtained. These data are all Windows API call requests made by the malware on the Windows 7 operating system.
	\item \textbf{Processing of Windows API calls}: We have observed 342 kinds of API calls in our dataset. These API calls are indexed with numbers 0-341 to create a new dataset. We have used the analysis results of the malware that had at least 10 different API calls in this data set.
	\item \textbf{Analysis of malware using Virus Total Public API} : In addition to our analyses, all malicious software in the data set was also analyzed by requesting the Virus Total service. In this way, each malware is studied by many different antivirus engines and their results are recorded.
	\item \textbf{Processing of analysis results}: The Virus Total service uses approximately 66 different antivirus applications for file analysis. Using the results of each study we obtained with this service, we identified each malware's families. As a result of our observations, we found that different antivirus applications for the same malicious software give different results. Besides, it was observed that not every antivirus application can detect some malicious software. For example; When the malware file with the hash value of 06e76cf96c7c7a3a138324516af9fce8 is analyzed in the Virus Total service, many applications indicate that this file is a worm, while some applications such as \textit{DrWeb} show that it is a trojan, and \textit{Babable} application indicates that this executable is a clean file. Therefore, while detecting each malware class, it is accepted that it belongs to the majority class of all analysis.
\end{enumerate}

\begin{figure}[h]
	\centering
	\includegraphics[width=1.0\linewidth]{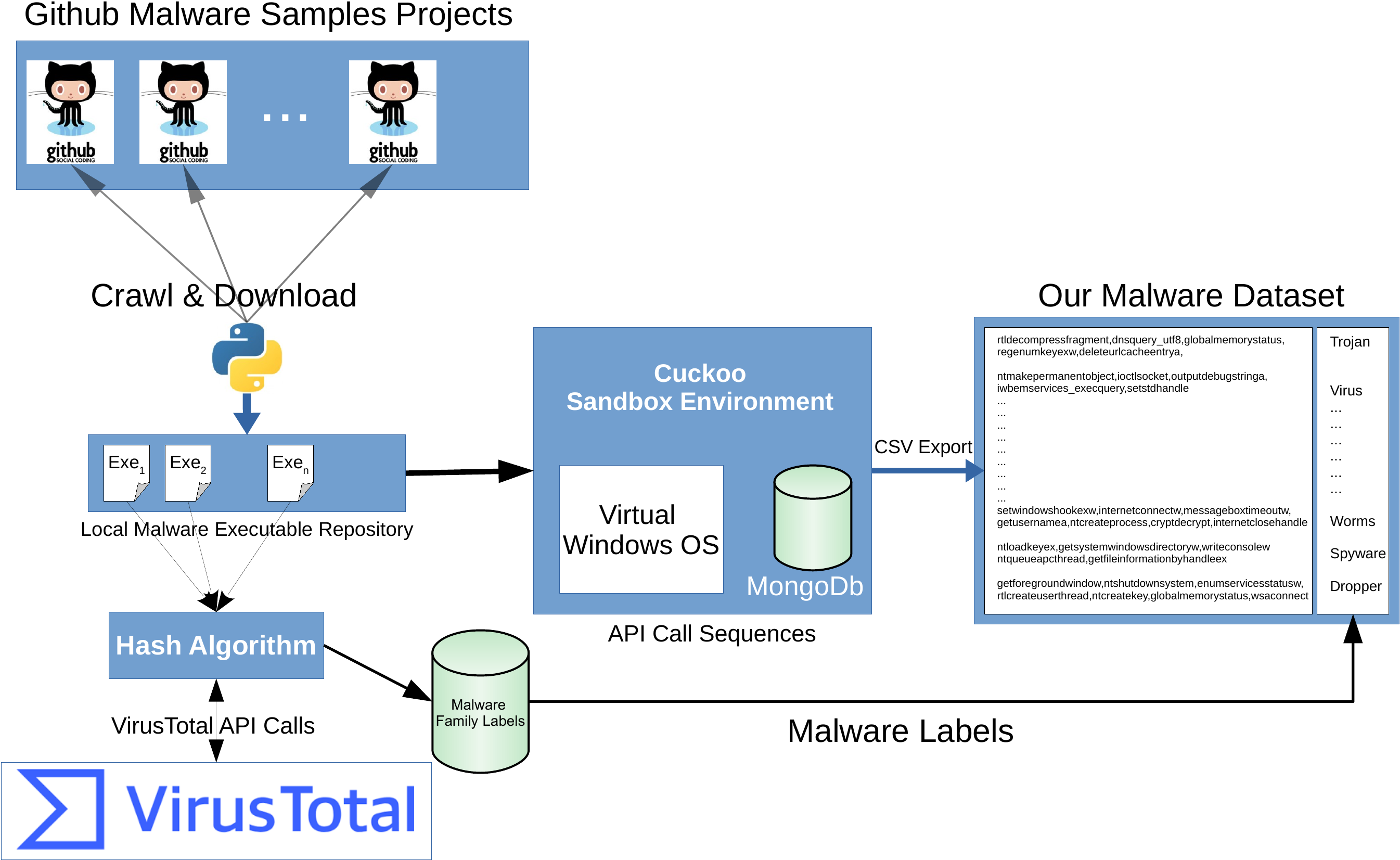}
	\caption{Overall representation of our data sources and technologies.}
	\label{fig:overall-data-collection}
\end{figure}

Figure \ref{fig:overall-data-collection} shows the general flow of the generation of the malware data set. As shown in the figure, we have obtained the MD5 hash values of the malware we collect from Github. We searched these hash values using the VirusTotal API, and we have received the families of these malicious software from the reports of 67 different antivirus software in VirusTotal. We have observed that the malicious software families found in the reports of these 67 other antivirus software in VirusTotal are different. 

Our research has translated the families produced by each of the software into 8 leading malware families: Trojan, Backdoor, Downloader, Worms, Spyware Adware, Dropper, Virus. Table \ref{tab:family} shows the number of malware belonging to malware families in our data set. As you can see in the table, the number of samples of other malware families except AdWare is quite close to each other. There is such a difference because we don't find too much malware from the adware malware family.

\begin{table}[H]
	\caption{Distribution of malicious software according to their families.}
	\label{tab:family}
	\centering
	\begin{tabular}{ccp{11.5cm}}
		\toprule
		\textbf{Malware Family} & \textbf{Samples} & \textbf{Description} \\ 
		\midrule
		Spyware & 832 & enables a user to obtain covert information about another's computer activities by transmitting data covertly from their hard drive. \\ 
		Downloader & 1001 & share the primary functionality of downloading content. \\ 
		Trojan & 1001 & misleads users of its true intent. \\ 
		Worms & 1001 & spreads copies of itself from computer to computer.  \\ 
		Adware & 379 & hides on your device and serves you advertisements.  \\ 
		Dropper & 891 &  surreptitiously carries viruses, back doors and other malicious software so they can be executed on the compromised machine. \\ 
		Virus & 1001 & designed to spread from host to host and has the ability to replicate itself.  \\ 
		Backdoor & 1001 & a technique in which a system security mechanism is bypassed undetectably to access a computer or its data. \\ 
		\midrule
		\textbf{Total} & 7107 & \\
		\bottomrule
	\end{tabular} 
\end{table}

\begin{minipage}[b]{0.5\linewidth}
		\includegraphics[width=1.0\linewidth]{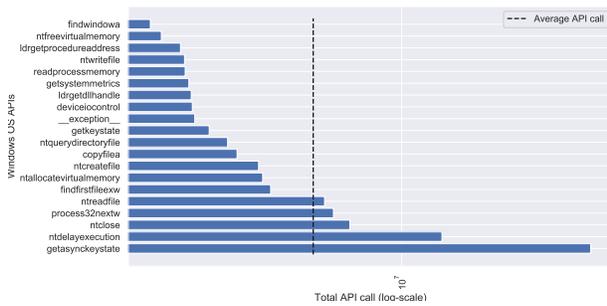}
		\captionof{figure}{The first 20 APIs used in our analysis.}
\end{minipage} \hfill
\begin{minipage}[b]{0.5\linewidth}
		\includegraphics[width=1.0\linewidth]{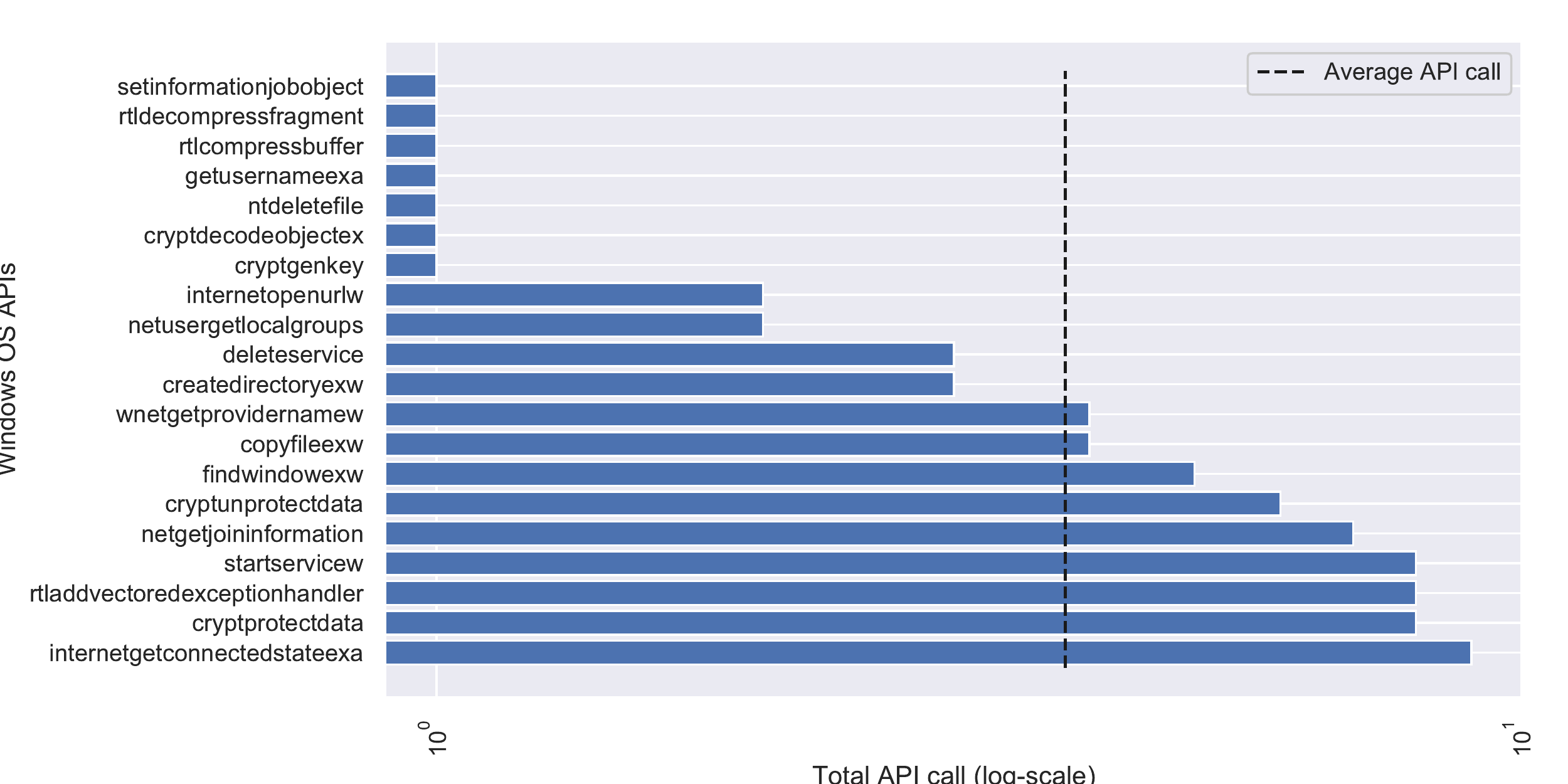}
		\captionof{figure}{The last 20 APIs used in our analysis.}
\end{minipage}

The operating system APIs used by malware families vary depending on their structure. These differences can be API or API calls. Malicious software can also make unnecessary API calls to hide and have a different signature on each client. The main reason for creating different signatures is that anti-viruses detect malicious software according to their signatures.

\begin{figure}[h]
	\centering
	\includegraphics[width=1.2\linewidth]{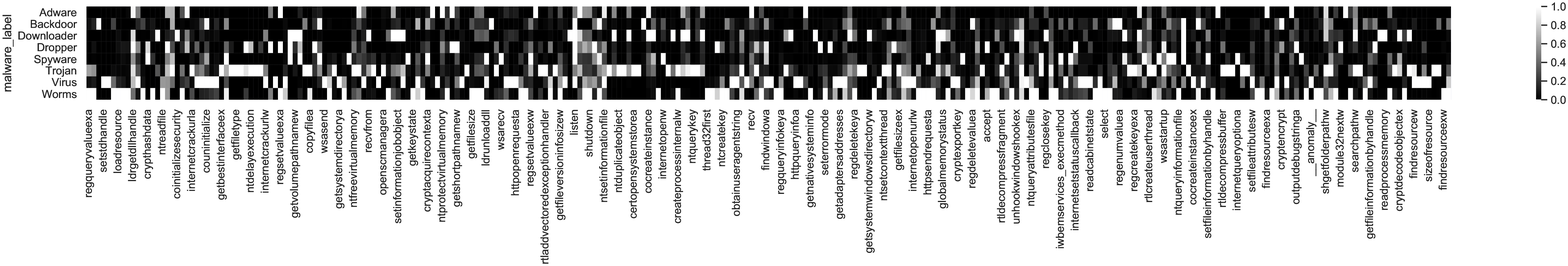}
	\caption{}
	\label{fig:malwaretypeheatmap}
\end{figure}

Figure \ref{fig:malwaretypeheatmap} shows the Windows operating system API calls' normalised values that belong to each family of malware.

\begin{figure}
	\begin{subfigure}[b]{0.45\textwidth}
		\includegraphics[width=\linewidth]{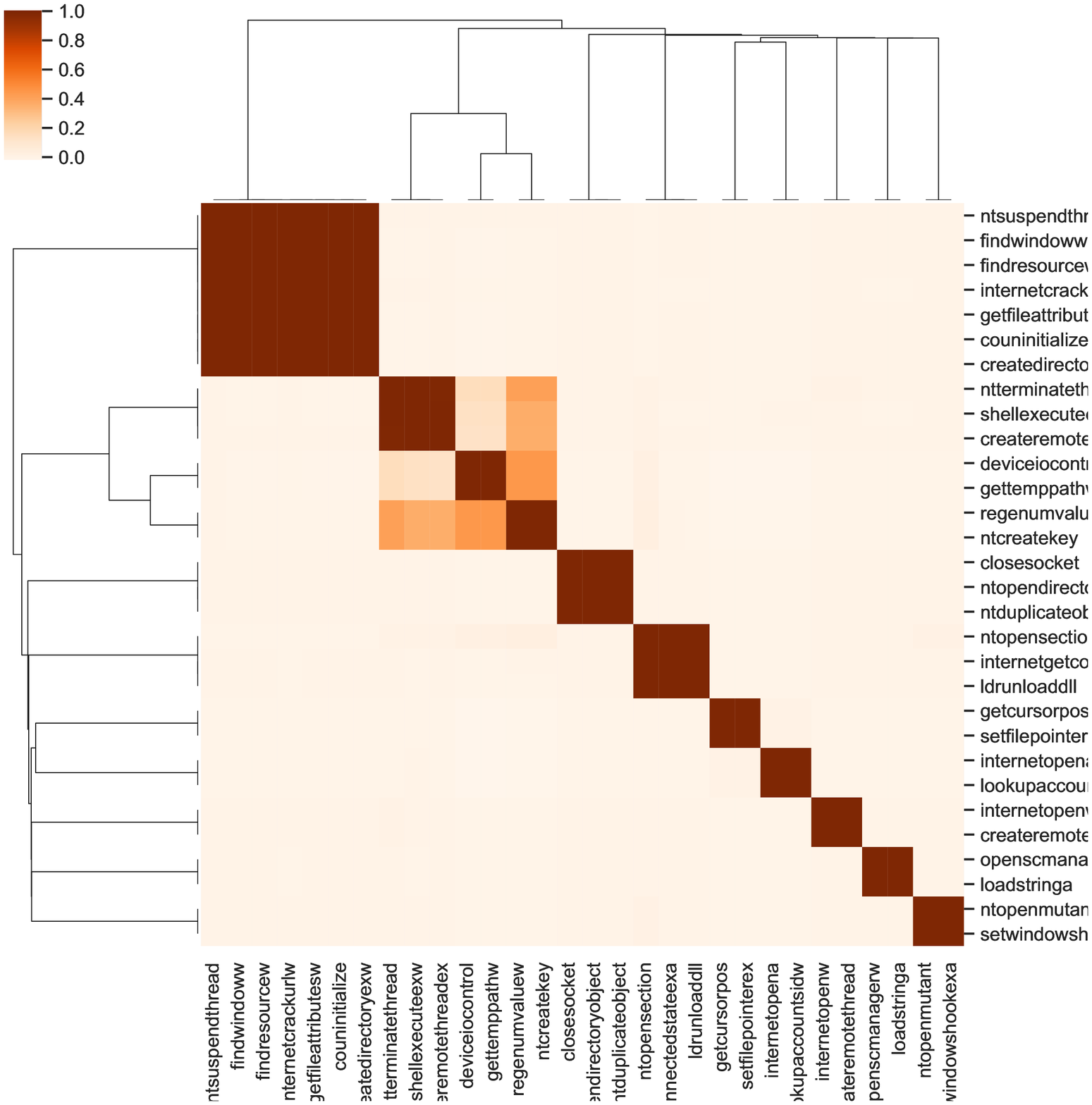}
		\caption{Downloader}
		\label{fig:downloader}
	\end{subfigure}%
	\begin{subfigure}[b]{0.45\textwidth}
		\includegraphics[width=\linewidth]{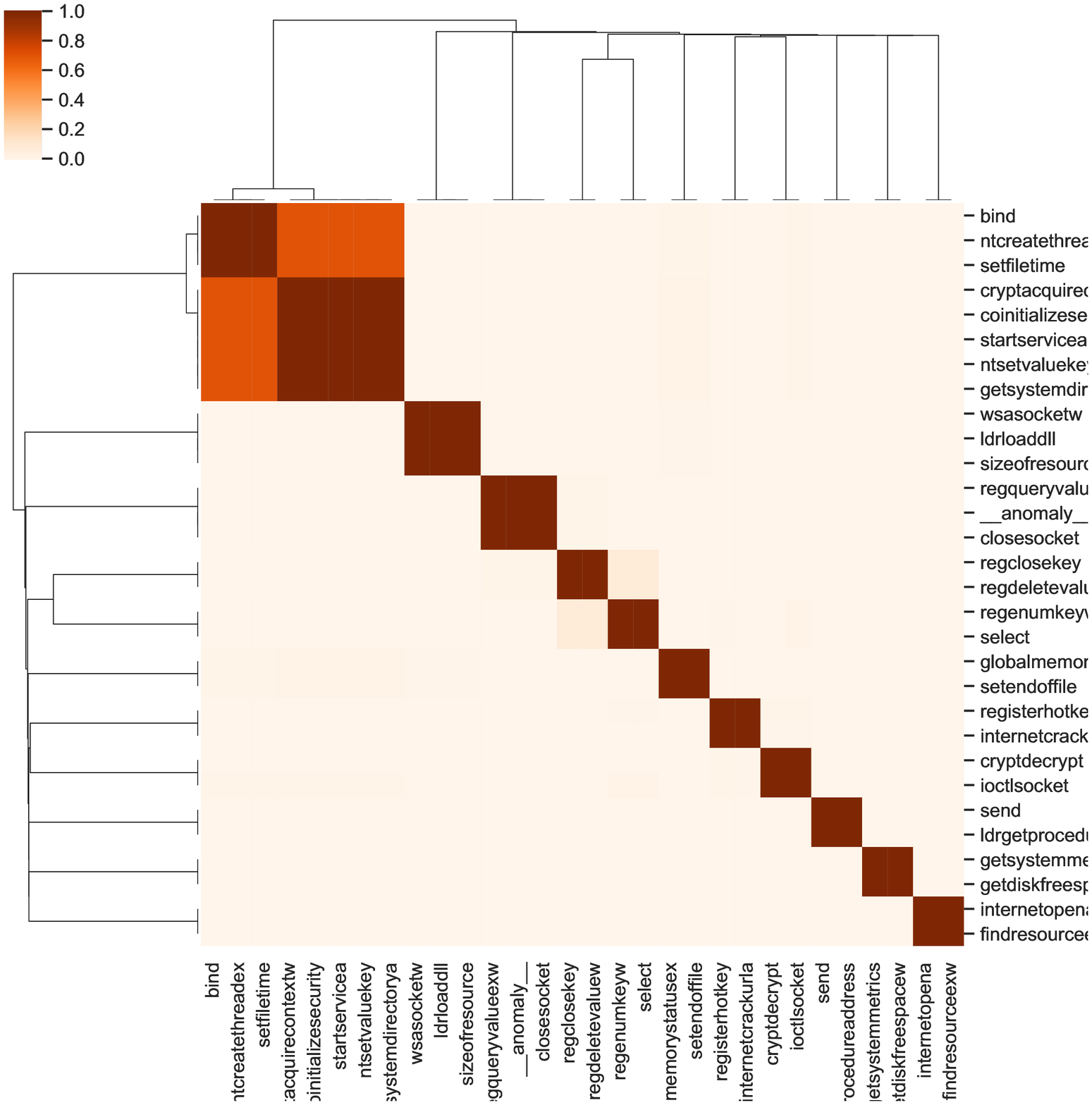}
		\caption{Worms}
		\label{fig:worms}
	\end{subfigure} 

	\begin{subfigure}[b]{0.45\textwidth}
		\includegraphics[width=\linewidth]{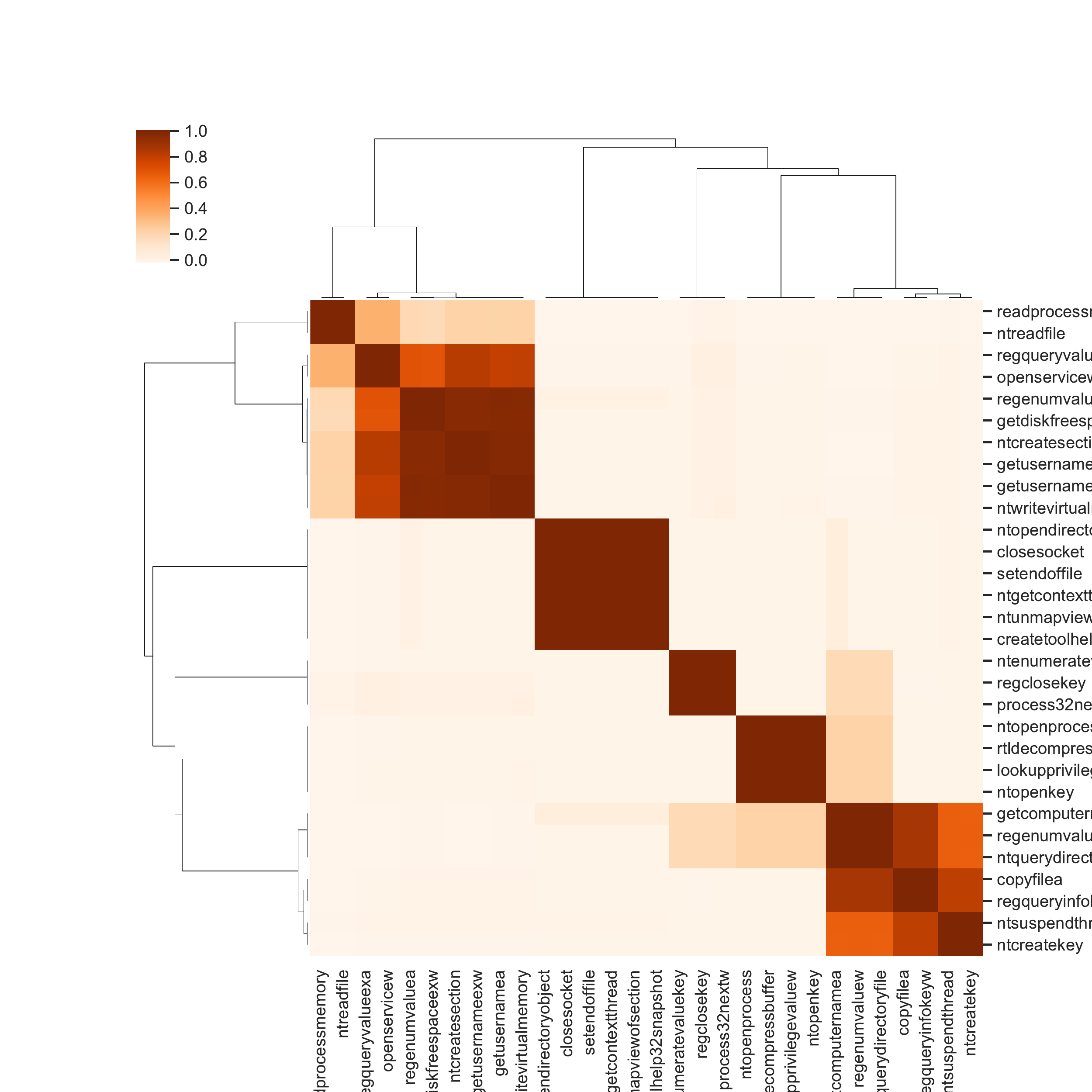}
		\caption{Spyware}
		\label{fig:Spyware}
	\end{subfigure}%
	\begin{subfigure}[b]{0.45\textwidth}
		\includegraphics[width=\linewidth]{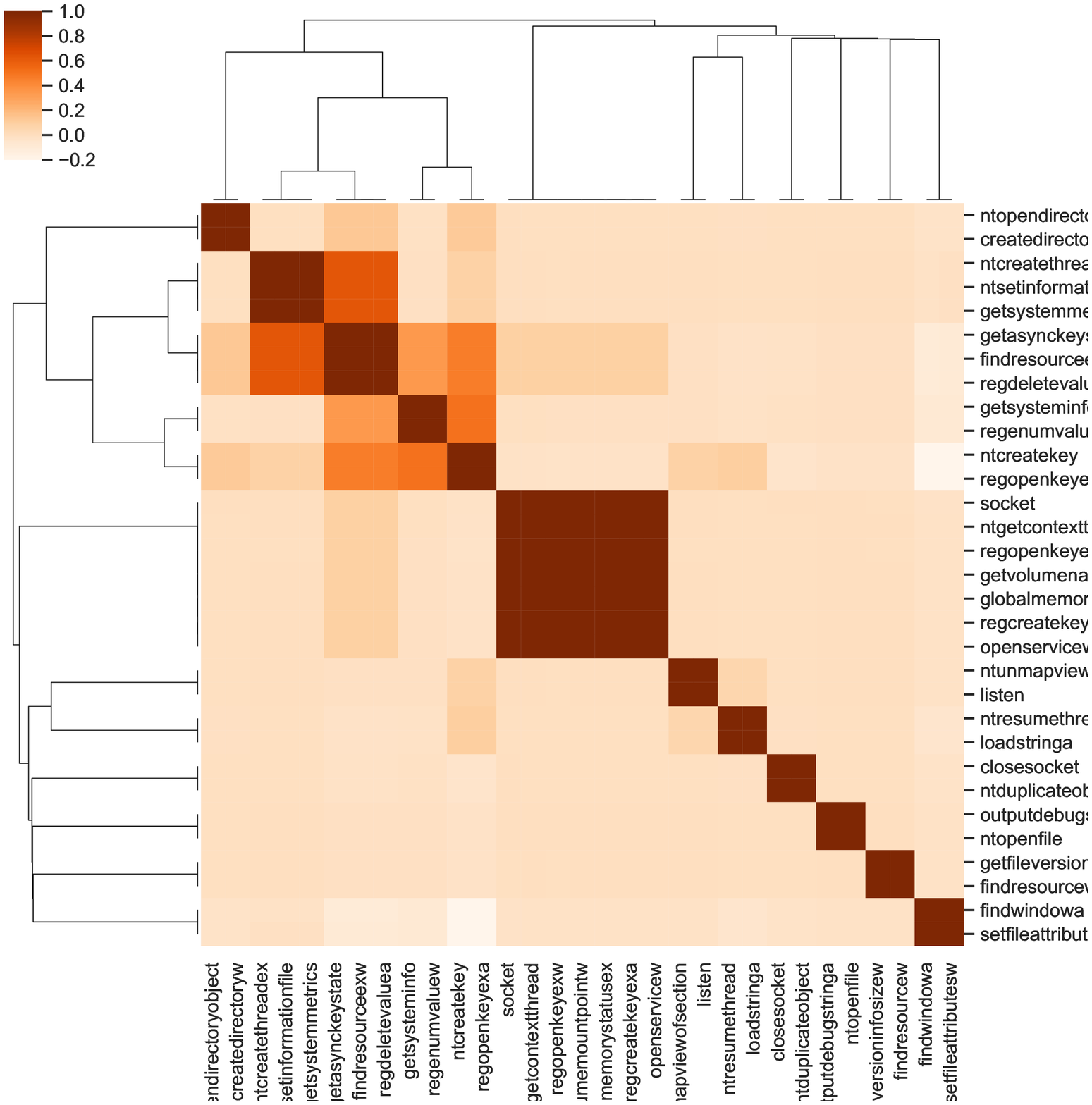}
		\caption{Adware}
		\label{fig:Adware}
	\end{subfigure}

	\begin{subfigure}[b]{0.45\textwidth}
		\includegraphics[width=\linewidth]{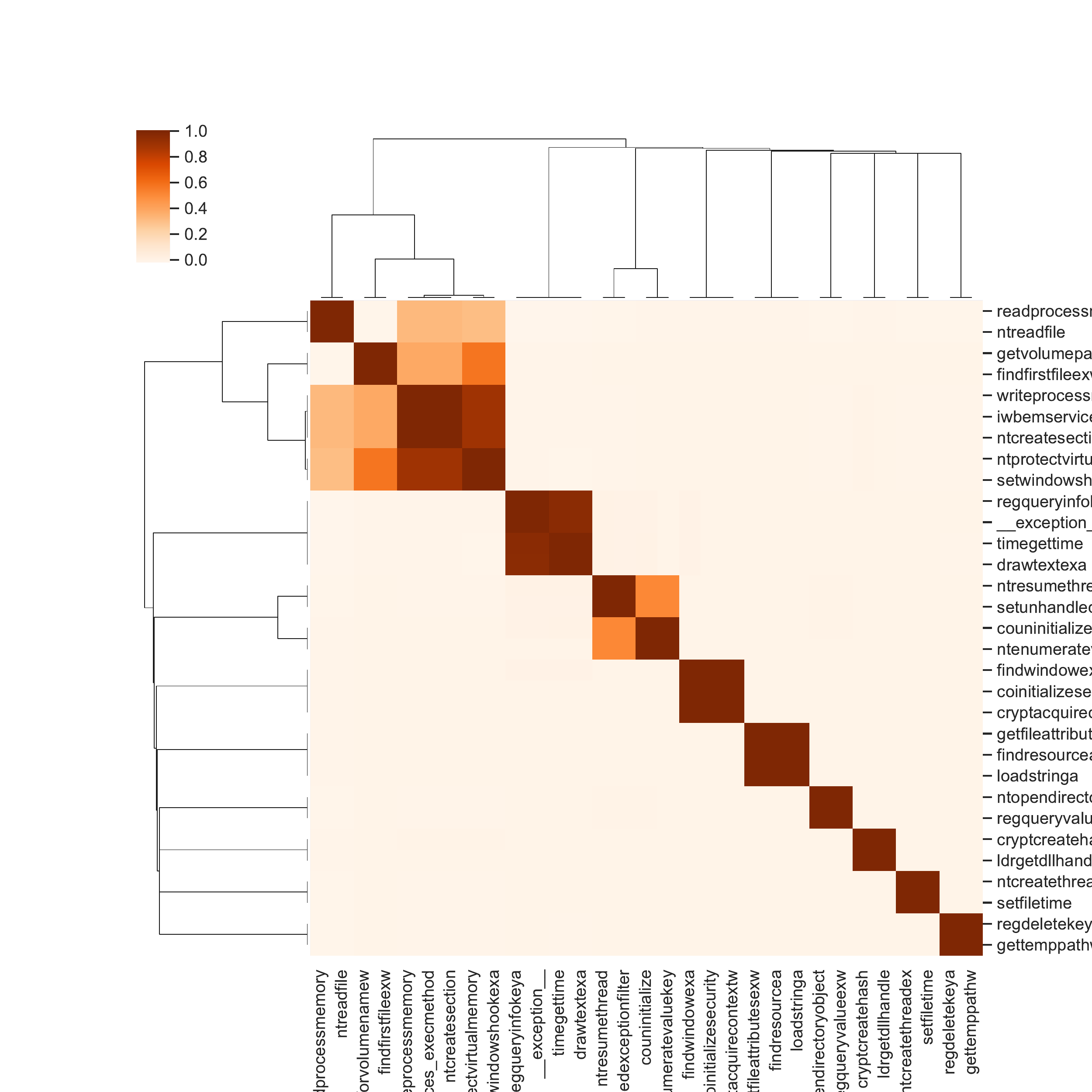}
		\caption{Dropper}
		\label{fig:Dropper}
	\end{subfigure}
	\begin{subfigure}[b]{0.45\textwidth}
		\includegraphics[width=\linewidth]{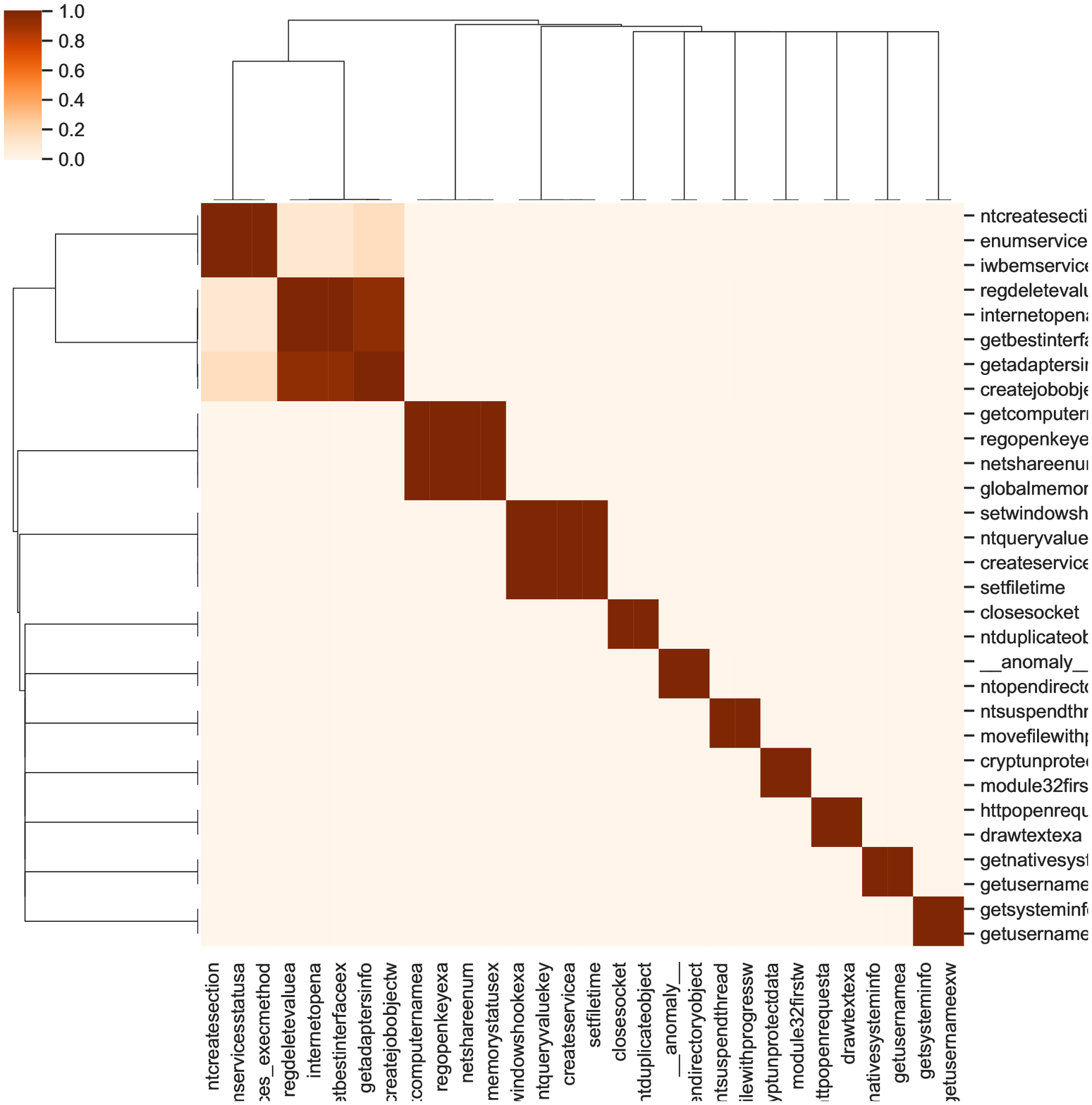}
		\caption{Virus}
		\label{fig:Virus}
	\end{subfigure}

	\caption{Windows API correlation heatmap.}\label{fig:corr-heatmap}
\end{figure}

Figure \ref{fig:corr-heatmap} shows the most correlated 30 API calls heatmap for each malware type. As can be seen from the figure, some APIs are called together for each family of malware. The malware follows the pre-defined API call sequence when performing its malicious activities. Although these API calls are different for different malware, they may be quite similar in malware families. Thus, the figure shows 5-10 APIs, positively correlated significantly for Dropper and Worms malware families.

\begin{table}[H]
	\caption{The most correlated API calls .}
	\label{tab:api-calls}
	\centering 
	\begin{tabular}{|p{2.3cm}c|p{2.3cm}c|p{2.3cm}c|p{2.6cm}c|} 
		\toprule
		\multicolumn{2}{c}{\textbf{Adware}} & \multicolumn{2}{c}{\textbf{Backdoor}} & \multicolumn{2}{c}{\textbf{Downloader}}  & \multicolumn{2}{c}{\textbf{Dropper}} \\ 
		\hline
		API Pairs & Corr & API Pairs & Corr & API Pairs & Corr & API Pairs & Corr \\
		\hline
		\scriptsize getfileversioninfosizew \newline findresourcew &  1.0 & \scriptsize regclosekey \newline process32nextw &  1.0 & \scriptsize ntopendirectoryobject \newline ntduplicateobject &  1.0 & \scriptsize iwbemservices\_execmethod \newline writeprocessmemory &  1.0 \\
		\hline
		\scriptsize regcreatekeyexa \newline openservicew &  1.0 & \scriptsize ntduplicateobject \newline ldrgetprocedureaddress &  1.0 & \scriptsize openscmanagerw \newline loadstringa &  1.0 & \scriptsize gettemppathw \newline regdeletekeya &  1.0 \\
		\hline
		\scriptsize findresourceexw \newline regdeletevaluea &  1.0 & \scriptsize internetopena \newline findresourceexw &  1.0 & \scriptsize getcursorpos \newline setfilepointerex &  1.0 & \scriptsize ntcreatethreadex \newline setfiletime &  1.0 \\
		\hline
		\scriptsize ntcreatethreadex \newline ntsetinformationfile &  1.0 & \scriptsize closesocket \newline ldrgetprocedureaddress &  1.0 & \scriptsize createremotethread \newline internetopenw &  1.0 & \scriptsize ntprotectvirtualmemory \newline setwindowshookexa &  1.0 \\
		\hline
		\scriptsize regdeletevaluea \newline getasynckeystate &  1.0 & \scriptsize getadaptersinfo \newline regqueryvalueexa &  1.0 & \scriptsize ntcreatekey \newline regenumvaluew &  1.0 & \scriptsize ntreadfile \newline readprocessmemory &  1.0 \\
		\hline
		\scriptsize ntcreatekey \newline regopenkeyexa &  1.0 & \scriptsize ntcreatethreadex \newline ntopenmutant &  1.0 & \scriptsize deviceiocontrol \newline gettemppathw &  1.0 & \scriptsize iwbemservices\_execmethod \newline ntcreatesection &  1.0 \\
		\hline
		\scriptsize getsystemmetrics \newline ntcreatethreadex &  1.0 & \scriptsize ntopenfile \newline unhookwindowshookex &  0.99 & \scriptsize closesocket \newline ntduplicateobject &  1.0 & \scriptsize ntcreatesection \newline writeprocessmemory &  1.0 \\
		\hline
		\scriptsize createdirectoryw \newline ntopendirectoryobject &  1.0 & \scriptsize ntclose \newline getbestinterfaceex &  0.99 & \scriptsize ntopendirectoryobject \newline closesocket &  1.0 & \scriptsize getvolumepathnamesforvol... \newline findfirstfileexw &  1.0 \\
		\hline
		\scriptsize findresourceexw \newline getasynckeystate &  1.0 & \scriptsize ntallocatevirtualmemory \newline getusernameexw &  0.99 & \scriptsize ntopenmutant \newline setwindowshookexa &  1.0 & \scriptsize ntopendirectoryobject \newline regqueryvalueexw &  1.0 \\
		\hline
		\scriptsize getsysteminfo \newline regenumvaluew &  1.0 & \scriptsize getsystemwindowsdir... \newline ntenumeratevaluekey &  0.99 & \scriptsize findresourcew \newline findwindoww &  0.99 & \scriptsize coinitializesecurity \newline cryptacquirecontextw &  0.99 \\
		\hline
		\scriptsize closesocket \newline ntduplicateobject &  1.0 & \scriptsize unhookwindowshookex \newline gethostbyname &  0.99 & \scriptsize getfileattributesw \newline findresourcew &  0.99 & \scriptsize findresourcea \newline loadstringa &  0.99 \\ 
		\hline \hline 
		\multicolumn{2}{c}{\textbf{Spyware}} & \multicolumn{2}{c}{\textbf{Trojan}} & \multicolumn{2}{c}{\textbf{Virus}}  & \multicolumn{2}{c}{\textbf{Worms}} \\
		\hline
		API Pairs & Corr & API Pairs & Corr & API Pairs & Corr & API Pairs & Corr \\
		\hline
		\scriptsize readprocessmemory \newline ntreadfile &  1.00 & \scriptsize  ntclose \newline createdirectoryw &  1.00 & \scriptsize closesocket \newline ntduplicateobject &  1.00 & \scriptsize internetopena \newline findresourceexw &  1.0 \\ 
		\hline
		\scriptsize regqueryvalueexa \newline openservicew &  1.00 & \scriptsize bind \newline wsasocketw &  1.00 & \scriptsize enumservicesstatusa \newline iwbemservices\_execmethod &  1.00 & \scriptsize getsystemdirectorya \newline coinitializesecurity &  1.00 \\
		\hline
		\scriptsize createtoolhelp32snapshot \newline ntgetcontextthread &  1.00 & \scriptsize getbestinterfaceex \newline regdeletevaluea &  1.00 & \scriptsize iwbemservices\_execmet. \newline ntcreatesection &  1.00 & \scriptsize startservicea \newline getsystemdirectorya &  1.00 \\
		\hline
		\scriptsize ntcreatesection \newline getusernameexw &  1.00 & \scriptsize ntcreatesection \newline setsockopt &  1.00 & \scriptsize getbestinterfaceex \newline regdeletevaluea &  1.00 & \scriptsize ntsetvaluekey \newline getsystemdirectorya &  1.00 \\
		\hline 
		\scriptsize createtoolhelp32snapshot \newline setendoffile &  1.00 & \scriptsize ntduplicateobject \newline ntcreatethreadex &  1.00 & \scriptsize getusernamea \newline getnativesysteminfo &  1.00 & \scriptsize ntcreatethreadex \newline setfiletime &  1.00 \\
		\hline
		\scriptsize ntopendirectoryobject \newline closesocket &  1.00 & \scriptsize searchpathw \newline createdirectoryw &  0.99 & \scriptsize internetopena \newline regdeletevaluea &  1.00 & \scriptsize startservicea \newline ntsetvaluekey &  1.00 \\
		\hline
		\scriptsize setendoffile \newline ntgetcontextthread &  1.00 & \scriptsize writeprocessmemory \newline createremotethread &  0.99 & \scriptsize enumservicesstatusa \newline ntcreatesection &  1.00 & \scriptsize closesocket \newline regqueryvalueexw &  1.00 \\
		\hline
		\scriptsize ntunmapviewofsection \newline ntgetcontextthread &  1.00 & \scriptsize regdeletevaluea \newline getusernameexw &  0.99 & \scriptsize setfiletime \newline createservicea &  1.00 & \scriptsize getsystemdirectorya \newline cryptacquirecontextw &  1.00 \\
		\hline
		\scriptsize createtoolhelp32snapshot \newline ntunmapviewofsection &  1.00 & \scriptsize getbestinterfaceex \newline getusernameexw &  0.99 & \scriptsize getadaptersinfo \newline createjobobjectw &  1.00 & \scriptsize setfiletime \newline bind &  1.0 \\
		\hline
		\scriptsize copyfilea \newline regqueryinfokeyw &  0.99 & \scriptsize httpsendrequestw \newline regqueryinfokeya &  0.99 & \scriptsize getadaptersinfo \newline createjobobjectw &  1.00 & \scriptsize ntcreatethreadex \newline bind &  1.00 \\
		\hline
	\end{tabular}
	
\end{table}

\bibliographystyle{unsrt}  
\bibliography{references}  

\end{document}